\documentclass[aps,twocolumn,showpacs,floatfix,superscriptaddress]{revtex4}
\usepackage{graphicx}
\usepackage{amsmath}
\usepackage{amsfonts}
\usepackage{amssymb}
\usepackage{epsfig}

\begin{document}

\newcommand{\la}{\langle}
\newcommand{\ra}{\rangle}
\def\be{\begin{equation}}
\def\ee{\end{equation}}
\def\bea{\begin{eqnarray}}
\def\eea{\end{eqnarray}}
\def\bma{\begin{mathletters}}
\def\ema{\end{mathletters}}
\newcommand{\one}{\mbox{$1 \hspace{-1.0mm}  {\bf l}$}}
\newcommand{\eins}{\mbox{$1 \hspace{-1.0mm}  {\bf l}$}}
\def\C{\hbox{$\mit I$\kern-.7em$\mit C$}}
\newcommand{\tr}{{\rm tr}}
\newcommand{\half}{\mbox{$\textstyle \frac{1}{2}$}}
\newcommand{\shalf}{\mbox{$\textstyle \frac{1}{\sqrt{2}}$}}
\newcommand{\ket}[1]{ | \, #1  \rangle}
\newcommand{\bra}[1]{ \langle #1 \,  |}
\newcommand{\proj}[1]{\ket{#1}\bra{#1}}
\newcommand{\kb}[2]{\ket{#1}\bra{#2}}
\newcommand{\bk}[2]{\langle \, #1 | \, #2 \rangle}
\def\II{I(\{p_k\},\{\rho_k\})}
\def\ss{{\cal K}}
\tolerance = 10000
\newcommand{\eps}\varepsilon




\title{Ground state cooling of atoms in optical lattices}
\author{M. Popp}
\affiliation{Max--Planck Institute for Quantum Optics,
Garching, Germany}
\author{J. J. Garcia-Ripoll}
\affiliation{Max--Planck Institute for Quantum Optics, Garching,
Germany}
\author{K. G. H.  Vollbrecht}
\affiliation{Max--Planck Institute for Quantum Optics, Garching,
Germany}
\author{J. I. Cirac}
\affiliation{Max--Planck Institute for Quantum Optics,
Garching, Germany}

\begin{abstract}
We propose two schemes for cooling bosonic and fermionic atoms
that are trapped in a deep optical lattice. The first scheme is a
quantum algorithm based on particle number filtering and state
dependent lattice shifts.  The second protocol alternates
filtering with a redistribution of particles by means of quantum
tunnelling. We provide a complete theoretical analysis of both
schemes and characterize the cooling efficiency in terms of the
entropy. Our schemes do not require addressing of single  lattice
sites and use a novel method, which is based on coherent laser
control, to perform very fast filtering.

\end{abstract}

\date{\today}
\pacs{03.75.Lm, 03.75.Hh, 03.75.Gg }

\maketitle
\section{Introduction} \label{Intro}
Ultracold atoms stored in optical lattices can be controlled and
manipulated with a very high degree of precision and flexibility.
This places them among the most promising candidates for
implementing quantum computations
\cite{OL_ent,L_OL_QC_03,W_OL_QC_02,Bloch03,VC04} and quantum
simulations of certain classes of quantum many--body systems
\cite{CZ_OL_review,Fermi-Hubbard,OLspin,JuanjoAKLT,Kagome,ringexchange,ZollerPM,LukinPM}.
However, both quantum simulation and quantum computation with this
system face a crucial problem: the temperature in current
experiments is too high. In this paper we propose and analyze two
methods to decrease the temperature and thus to reach the
conditions required to observe the interesting regimes in quantum
simulations and quantum computation.

So far, several experimental groups have been able to load bosonic
or fermionic atoms in optical lattices and reach the strong
interaction regime
\cite{Bloch02,Tonks,ETH_Tonks,ETH_Fermi,NIST_OL_04,MIT_OL_Bose,Texas_OL05,Grimm_OL_06}.
 In those experiments, the typical temperatures are still
relatively high. For instance, the analysis of experiments in the
Tonks gas regime indicates a temperature of the order of the width
of the lowest Bloch band \cite{Tonks}, and for a Mott Insulator
 a temperature of the order of the
on-site interaction energy has  been reported
\cite{temp_Mott,ETH_Tonks}. For fermions one observes temperatures
of the order of the Fermi energy \cite{temp_Fermi, temp_Fermi_ETH,
ETH_Fermi_mol}. Those temperatures put strong restrictions on the
physical phenomena that can be observed with those systems and
also on the quantum information tasks that can be carried out with
them. They stem from the fact that atoms are loaded adiabatically
starting from a Bose--Einstein condensate (in the case of bosons).
On the one hand, the original condensate has a relatively high
entropy \cite{BECStringari} that is inherited by the atoms in the
lattice in the adiabatic process. On the other hand, the process
may not be completely adiabatic, which gives rise to heating.
Thus, it seems that the only way of overcoming these problems is
to cool the atoms once they have been loaded in the optical
lattice.

One may think of several ways of cooling atoms in optical
lattices. For example, one may use sympathetic cooling with a
different Bose--Einstein condensate \cite{Fermi-Hubbard,DFZ04}.
Here we propose two alternative schemes which do not require the
addition of a condensate. They aim at cooling  atoms to the
 ground state of the Mott-insulating (MI) regime and allow us to predict
  temperatures which are low enough for
practical interests. Our protocols are based on translation
invariant operations (i.e. do not require single--site addressing)
and include the presence of an additional harmonic trapping
potential, as it is the case in present experiments. Although we
will be mostly analyzing their effects on bosonic atoms, they can
also be used for fermions.

The first scheme is based on the repeated application of
occupation number filtering \cite{RZ03}. Via tunnelling, particles
from the borders of the trap are transferred to the center, where
they are discarded by subsequent filter operations.
  The second scheme combines   filtering
 with algorithms inspired by quantum computation
\cite{VC04} and hence will be termed
 \emph{algorithmic cooling of atoms}
\cite{NMR}. The central idea is to split the atomic cloud into two
components and to use particles at the border of one  component as
"pointers" that remove "hot" particles at the borders of the other
component. We provide a detailed theoretical description of our
cooling schemes and compare our theory with exact numerical
calculations. In particular, we quantify the cooling efficiency
analytically in terms of the initial and final entropy. We find
that filtering becomes more efficient at low temperatures. This
feature makes it possible to reach states very close to the ground
state after only a few subsequent filtering operations. Our theory
further predicts that the algorithmic protocol is more efficient
at higher temperatures and that the final entropy per particle
becomes zero  in the thermodynamic limit. In addition,
experimental requirements and time scales are discussed.

Since filtering is an important ingredient of all our protocols,
we have devised an fast physical implementation  which is based on
optimal coherent laser control. Already  comparatively simple
optimization schemes work on a time scale that is significantly
shorter  than the one in \cite{RZ03} and \cite{Bloch_filter06}.

The paper is organized as follows. We start in Sect. \ref{Init}
with reviewing the physical system in terms of the Bose-Hubbard
model and discussing  realistic initial state variables such as
entropy and particle number. In Sect. \ref{Sect_filtering} we give
a detailed
 theoretical analysis of filtering under realistic experimental conditions.
Building on this,  we study in Sect. \ref{Sect_repeatF} the
repeated application of filtering. An algorithmic  ground state
cooling protocol is proposed and analyzed in
 Sect. \ref{Sect_gscool}. Next,  Sect. \ref{Sect_discussion} is dedicated to the
 discussion of our protocols, including a comparison of analytical results with
 numerical calculations. A further central result of our work is presented in
 Sect. \ref{Sect_ultrafast}. There we propose
  an ultra-fast
implementation of filtering operations based on coherent laser
 control. We conclude with some remarks concerning possible variants and
 extensions of our protocols. In the Appendix we develop a
 fermionization procedure of the Bose-Hubbard Hamiltonian,
 which accounts for up to two particles per site.

\section{Initial state and basic concepts} \label{Init}
We consider  a gas of ultra-cold bosonic atoms which have been
loaded into a three dimensional (3D) optical lattice.  The lattice
depth is proportional to the dynamic atomic polarizability times
the laser intensity. We further account for an additional harmonic
trapping potential which either arises naturally from the Gaussian
density profile of the laser beams or can be controlled separately
via an external magnetic or optical confinement
\cite{Tonks,Bloch05}.

In the following we will  restrict ourselves to one-dimensional
(1D) lattices, i.e. we assume that tunnelling is switched off for
all times along the transversal lattice directions. This system is
most conveniently described in terms of a single band Bose-Hubbard
model \cite{oplat98}. For a  lattice of length $L$ the Hamiltonian
in second quantized form reads {\small \be \label{BHM}
 H_{BH}=\sum_{k=-L/2}^{L/2} \left[-J (a^\dagger_k a_{k+1} + h.c.)+
  \frac{U}{2} n_k (n_k-1)+ b  k^2 n_k \right] .
\ee} The parameter $J$ denotes the hopping matrix element between
two adjacent sites, $U$ is the on-site interaction energy between
two atoms and the energy $b$ accounts for the strength of the
harmonic confinement. Operators $a_k^\dagger$ and $a_k$ create and
annihilate, respectively, a particle on site $k$, and
$n_k=a^\dagger_k a_k$ is the occupation number operator. When
raising the laser intensity the hopping rate decreases
exponentially, whereas the interaction parameter $U$ stays almost
constant \cite{oplat98}. Therefore we have adopted $U$ as  the
natural energy unit of the system.

In the following we will consider 1D thermal states in the grand
canonical ensemble, which are characterized by two additional
parameters, temperature $k T = 1/ \beta$ and chemical potential
$\mu$. In particular, we are interested in the no-tunnelling limit
\cite{comment_no-tunnel}, $J\rightarrow 0$, in which the
Hamiltonian (\ref{BHM}) becomes diagonal in the Fock basis of
independent lattice sites: $\{ |n_{-(L-1)/2} \ldots n_0 \ldots
n_{(L-1)/2}\ra \}$. The density matrix then factorizes into a
tensor product of thermal states for each lattice site: \be
\label{rho_ind} \rho=\frac{1}{\Theta} e^{-\beta (H_{BH}- \mu
  N)}=\bigotimes_{k=-L/2}^{L/2} \rho_k, \ee which  simplifies
  calculations considerably.
For  instance,
 the von-Neumann entropy can be written as,
\be S(\rho)=\tr( \rho \log_2 \rho)=\sum_k
S(\rho_k).\label{vonNeumann}\ee This quantity will be
central in this article, because it allows to assess the
cooling performance of our protocols. To be more precise,
we define two figures of merit. The ratio of the entropies
\emph{per} particle after and before invoking the protocol,
$s_f/s_i$, quantifies the amount of cooling. The ratio of
the final and initial number of particles, $N_f/N_i$,
quantifies the particle loss induced by the protocol.

Note, however, that the entropy $S(\rho_{f})$ is only a good
figure of merit if the state  $\rho_f$ after the cooling protocol
 is close to thermal equilibrium. If this is not
fulfilled, we compute an effective thermal state, $ \rho_f
\rightarrow \rho_\textrm{eff}$, by accounting for particle number
and energy conservation in closed, isolated systems. This is
performed numerically by tuning the chemical potential and
temperature of a thermal state $\rho_\textrm{eff}$ until the
expectation values for particle number and energy coincide with
the ones of the original state $\rho_f$.  This procedure can be
implemented rather easily in the no-tunnelling regime, in which
the density matrix factorizes (\ref{rho_ind}).

Our figures of merit can then be calculated from
$\rho_\textrm{eff}$. For instance, the final entropy is given by
$S(\rho_\textrm{eff})$. It constitutes the maximum entropy of a
state which yields the same expectation values for energy $E$ and
particle number $N$ as the final state $\rho_f$. In this context
it is important to point out that other variables, like energy or
temperature, are not very well suited as figures of merit, because
they depend crucially on external system parameters such as  the
harmonic trap strength.

We now study the structure of the initial state in more detail. To
this end we first give typical parameter values.  The analysis of
recent experiments in the MI regime implies a substantial
temperature of the order of the on-site interaction energy
\cite{Tonks,ETH_Tonks,temp_Mott}. This result is consistent with
our own numerical calculations \cite{Cool2} and translates into an
entropy per particle $s:= S/N \approx 1$. The particle number in a
1D tube of a 3D lattice as in \cite{Bloch02} typically ranges
between $N=10$ and $N=130$ particles. A representative density
distribution corresponding to such initial conditions (with
$N=65$) is plotted in Fig. \ref{fig_P1}a. In this example the
inverse temperature is given by $\beta U=4.5$. Since our cooling
protocols lead to even lower temperatures,  we will from now on
focus on the \emph{low temperature regime}, $\beta U \gg 1$.
Moreover, we will only consider states with at most  two particles
per site, which puts the constraint $\mu \lesssim 2 U- 1 / \beta $
on the chemical potential. Such a situation can either be achieved
by choosing the harmonic trap shallow enough or by applying an
appropriate filtering operation \cite{RZ03}.

Under  the assumptions $e^{\beta U} \gg 1$ and $\mu -U/2 \gtrsim b
+ 1/(2 \beta)$  we will now show that the  density distribution of
the initial state (\ref{rho_ind}) can be separated into regions
that are completely characterized by fermionic distribution
functions of the form: \be \label{nf} f_k (b, \beta, \mu) =
\frac{1}{1+e^{ \beta (b k^2 -\mu)}}. \ee To be more precise, for
sites at the borders of the density distribution, $b k^2 \gg
\mu-U/2 +1 /(2 \beta)$, the mean occupation number is given by
$\la n_k \ra \approx n_\textrm{I}(k)$ with
$n_\textrm{I}(k):=f_k(b, \beta, \mu)$. In the center of the trap,
$b k^2 \ll \mu-U/2 -1 /(2 \beta)$, one has: $\la n_k \ra \approx
1+ n_\textrm{II}(k)$ with $n_\textrm{II}(k):=f_k(b, \beta,
\mu_\textrm{II})$ and effective chemical potential
$\mu_\textrm{II}:=\mu-U$ (see e.g. Fig.~\ref{fig_P1}a).

 Starting from state (\ref{rho_ind}), with parameters $\beta$, $\mu$ and $b$, the
grand canonical partition function for site $k$ is given by: \be
\label{Theta_k} \Theta_k = 1 + a ~x_k +b ~x_k^2,\ee with
$x_k=e^{-\beta b k^2}$, $a=e^{\beta \mu}$ and $b=e^{\beta (2
\mu-U)}$. In this notation the probabilities $p_k^n$ of finding $n$
particles at site $k$ can be written as: $p_k^0=1/\Theta_k$,
$p_k^1=a~x/\Theta_k$ and $p_k^2=b~x^2/\Theta_k$. For analyzing these
functions we split the lattice into a central region and two border
regions. For lattice sites at the borders one finds $b x^2 \ll 1,
ax$, meaning that the probability for doubly occupied sites becomes
negligible: $p_k^2 \ll p_k^0, p_k^1$. The average occupation is thus
given by $\la n_k \ra \approx p_k^1$, with \be p_k^1 \approx \frac{a
x}{1 +ax }= \frac{1}{1+e^{\beta (b k^2-\mu)}} . \ee In the crossover
region, $ b k^2 \approx \mu-U/2$, one obtains a MI phase
($p_k^0,p_k^2 \ll p_k^1\approx 1$). In the center of the trap one
finds a negligible probability for empty sites: $p_k^0 \ll p_k^1,
p_k^2$, since $ax, bx^2 \gg 1$. The average population at site $k$
becomes $\la n_k \ra = p_k^1+ 2~p_k^2 \approx 1 + p_k^2$, where \be
p_k^2 \approx \frac{b x^2}{a x + b x^2}= \frac{1}{1+e^{\beta ( b k^2
- (\mu-U))}}. \ee This is identical to the fermionic distribution
(\ref{nf}) with effective chemical potential $\mu-U$. Hence the
density distribution in this lattice region can be interpreted as a
thermal distribution of hard-core bosons (phase II in
Fig.~\ref{fig_P1}a) sitting on top of a MI phase with unit filling.
Note that this central MI phase is well reproduced by the function
$n_\textrm{I}(k)$ , which originally has been derived for the border
region. As a consequence, the density distribution for the whole
lattice can be put in the simple form: $\la n_k \ra \approx
n_\textrm{I}(k)+n_\textrm{II}(k)$, which corresponds to two
fermionic phases I and II [Fig. \ref{fig_P1}a], sitting on top of
each other. In other words, the initial state of our system can
effectively be described in terms of non-interacting fermions, which
can occupy two different energy bands I and II, with dispersion
relations $\eps_\textrm{I}=b k^2$ and $\eps_\textrm{II}=b k^2  +U$,
respectively [see also Appendix A and \cite{PC05}].

The initial density profile can be further  characterized by two
distinctive points. At sites
 $k=\pm k_\mu := \sqrt{\mu/b}$, which correspond to the Fermi levels
of phase I,  one obtains $\la n_{k_\mu} \ra=1/2$. Hence, $k_\mu $
determines the radius of the atomic cloud. Note also that in the
case $\mu \approx U$ singly occupied sites around the Fermi levels
become degenerate with doubly occupied sites at the center of the
trap. At the central site ($k=0$) one one finds an average
occupation: \be \la n_0 \ra=1+ \frac{1}{1+e^{\beta (U-\mu)}}. \ee
For instance, the value $\la n_0 \ra=3/2$ fixes the chemical
potential to be $\mu=U$.

\section{Analysis of Filtering} \label{Sect_filtering}
By filtering we denote the state selective  removal of atoms from
the system, depending on the single site occupation number
\cite{RZ03}. For instance, this can be achieved with a unitary
operation \be \label{U_nm} U_{m,0}^{M,m-M}: |m,0\rangle
\leftrightarrow |M,m-M\rangle, \ee that transfers $m-M$ particles
from a  Fock state with $m$  atoms in internal states $|a\ra$ to
an initially unoccupied level $|b\ra$. Particles in this level are
removed subsequently and the process is repeated for all $m>M$.
The maximum single site occupation number then becomes $M$.
Alternatively, filtering can be described in terms of a completely
positive map acting solely on the density operator of atoms
$|a\ra$:
\begin{eqnarray}
\label{Fn}
F_M &:& \sum_{n,m} \rho_{n,m}|n\rangle\langle m| \to\nonumber\\
&&\to\sum_{n,m\leq M}\rho_{n,m}|n\rangle\langle m| + \sum_{n>
M}\rho_{n,n} |M\rangle\langle M|.
\end{eqnarray}
In particular, we are interested in the filtering operation $F_1$
which yields either empty or singly occupied sites. This operation
can be carried out with the scheme introduced in \cite{RZ03},
which is based on the blockade mechanism due to atom--atom
interactions. It enables a state selective adiabatic transfer of
particles from one internal state of the atom to another.
Recently, an alternative scheme which relies on resonant control
of  interaction driven spin oscillations has been put forward
\cite{Bloch_filter06}. However, the predicted operation times of
both approaches are comparatively long. Since fast filtering is
crucial for the experimental realization of our protocols, we will
propose in Sect. \ref{Sect_ultrafast} an ultra-fast, coherent
implementation of $F_1$, relying on optimal laser control.

\begin{figure}[h]
\centering \epsfig{file=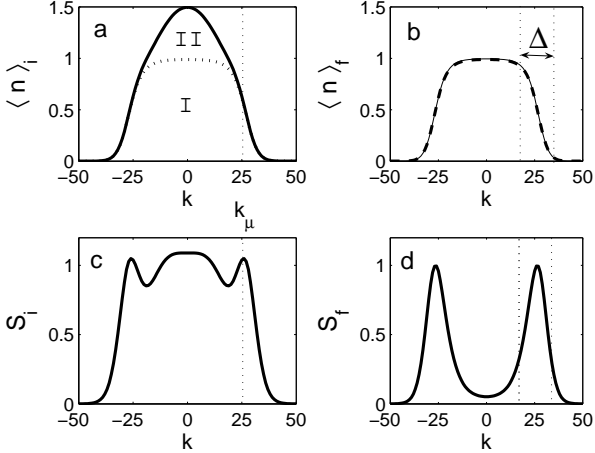,width=\linewidth} \caption{Spatial
dependence of the average particle number $\la n \ra$ and entropy
$S$ before (left) and after (right) the application of the filter
operation $F_1$. The final particle distribution can be well
described by Eq. (\ref{nf}) (dashed line in (b)). Numerical
parameters for initial state: $N_i=65$, $s_i$=1 and $U/b=700$
($\beta U=4.5$, $\mu/U=1$). Figures of merit: $s_f/s_i=0.56$ and
$N_f/N_i=0.80$.} \label{fig_P1}
\end{figure}
In Fig.~\ref{fig_P1} we study the action of $F_1$ on a realistic
1D thermal state in the no-tunnelling regime, as defined in the
previous section. One observes that a nearly perfect MI phase with
filling factor $\nu=1$ is created in the center of the trap.
Defects in this phase are due to the presence of holes and
preferably locate at the borders of the trap. This behavior is
reminiscent of fermions, for which excitations can only be created
within an energy range of order $kT$ around the Fermi level. This
numerical observation can easily be understood with our previous
analysis of the initial state. Filtering removes phase II, which
is due to doubly occupied sites, and leaves the fermionic phase I
unaffected [Fig. \ref{fig_P1}a].

Let us now study the cooling efficiency of  operation $F_1$. This
means we have to compute the entropy per particle of the states
before and after filtering. According to our preceding discussion
this problem reduces to the computation of the entropy $S$ and the
particle number $N$, corresponding to the bands  I and II. The
entropy of a fermionic distribution of the form
  (\ref{nf})
 is given by: \bea \label{S_fermi}
  S_F(b, \beta, z)&=& \frac{1}{\sqrt{\beta b}}~  \left[ \sigma (z)\right.
 + \frac{\sqrt{\pi}}{2 \ln 2} ( 2 ~ \ln z ~ \textrm{Li}_{1/2}(-z) \nonumber \\
  &-& \left.\textrm{Li}_{3/2}(-z) ) \right], \eea
with fugacity $z=e^{\beta \mu}$ and $\textrm{Li}_{n}(x)$ denoting
the polylogarithm functions.
 The
 function $\sigma(z)$ is defined as the integral
 { \small \be \label{sigma_f}
 \sigma(z):=  \int_{-\infty}^\infty dx  ~ \log_2 \left( 1+z ~e^{-x^2} \right)
  .
 \ee}
 For phase I one can find a simpler expression for the entropy (\ref{S_fermi})  by
 expansion  around the Fermi level $k=k_\mu+ dk$. Note that the range of validity, $|dk| \ll 1/\sqrt{\beta b}$,
 of this approximation covers  all lattice sites that give a significant contribution to the total entropy.
  This yields the following relations:
\be S_\textrm{I} \approx   \sigma_\textrm{I}  \frac{2}{\beta
\sqrt{b \mu}}, \quad
 N_\textrm{I} \approx  2 \sqrt{\frac{\mu}{b}}  \label{S_I}, \ee
 with  $\sigma_\textrm{I}:=\pi^2/(6 \ln 2)$.
  For phase II this approach is typically not valid and one   obtains the
  general expressions:
 \be S_\textrm{II}= S_F(b,\beta,z_\textrm{II}), \quad N_\textrm{II}=- \sqrt{\frac{\pi}{\beta b}}~ \textrm{Li}_{1/2}(z_\textrm{II}),\ee
  with $z_\textrm{II}=e^{\beta
  \mu_\textrm{II}}$.
 In the special case \mbox{$\mu=U$}
\cite{com_optimal} one can simplify the above expressions to: \be
 S_\textrm{II} \approx  \sigma_\textrm{II}  \frac{1}{\sqrt{\beta
 b}}, \quad
  N_\textrm{II} \approx  \eta_\textrm{II} \frac{1}{\sqrt{\beta b}}, \label{S_II} \ee
with numerical coefficients $\sigma_\textrm{II} \approx 2.935$ and
$\eta_\textrm{II}=(1-\sqrt{2}) \sqrt{\pi} \zeta (1/2)\approx
1.063$.

With these findings we can now give a quantitative interpretation
of  Fig. \ref{fig_P1}. The initial entropy is composed of two
components: $S_i=S_\textrm{I}+S_\textrm{II}$. Filtering removes
the contribution $S_\textrm{II}$, which arises from the
coexistence of singly and doubly occupied sites. The final entropy
is thus given by $S_f=S_\textrm{I}$. This residual entropy is
localized around the Fermi levels $-k_\mu$ and $k_\mu$ within a
region of width $\Delta$ [Fig.~\ref{fig_P1}]: \be \label{Delta}
\Delta= \frac{2}{\beta \sqrt{b \mu}}=  \frac{N_\textrm{I}}{\beta
\mu}, \ee and one can write $S_f=\sigma_\textrm{I} \ \Delta$. For
the initial and final particle numbers one has the corresponding
relations: $N_i=N_\textrm{I}+N_\textrm{II}$ and
$N_f=N_\textrm{I}$. Hence, the final entropy per particle can be
written as: \be \label{s_f} s_f=\frac{S_f}{N_f}\approx
\sigma_\textrm{I} \frac{1}{\beta \mu}. \ee For the special choice
$\mu=U$ (or equivalently \mbox{$\la n_0 \ra=1.5$}) one finds the
following expressions for our figures of merit: \bea
 \frac{s_f}{s_i}&\approx& \frac{\sigma_\textrm{I}}{
 \sqrt{\beta U}} \frac{\eta_\textrm{II}+2 \sqrt{\beta
U}}{\sigma_\textrm{II} \sqrt{\beta U} +
2~\sigma_\textrm{I}}, \label{cool_eff} \\
\frac{N_f}{N_i} & \approx & \frac{1}{1+ \frac{\eta_\textrm{II}}{2
\sqrt{\beta U}}}. \eea   This result shows that filtering becomes
more efficient with decreasing temperature, since $s_f/s_i \propto
1/\sqrt{\beta U}\rightarrow 0$ and $N_f/N_i \rightarrow 1$ for
$\beta U \rightarrow \infty$.

It is important to note that the state after filtering is not an
equilibrium state, because it is energetically favorable that
particles tunnel from the borders to the center of the trap,
thereby forming doubly occupied sites. According to the discussion
in the previous section this implies that the final entropy, which
enters the cooling efficiency, should  be calculated from an
effective state $\rho_{eff}$ after equilibration. However, this
would yield a rather pessimistic estimate for the cooling
efficiency. Since the final density $n_\textrm{I}(k)$ already has
the form of a  thermal distribution function, one can easily come
up with a much simpler (and faster) way to reach a
thermodynamically stable configuration. While tunnelling is still
suppressed, one has to decrease the strength of the harmonic
confinement to a new value $b'$, with $b' \leq b ~ U/(2 \mu)$. The
system is then in the equilibrium configuration $f_k(b', \beta',
\mu')$ (\ref{nf}) with rescaled parameters $T'=T~b'/b$ and
$\mu'=\mu~b'/b \leq U/2$. This observation shows that it is
misleading to infer the cooling efficiency solely from the ratio
$T'/T$, because it depends crucially on the  choice of $b'$. Note
also that this procedure indeed  allows to achieve the predicted
value (\ref{cool_eff}) for the cooling efficiency.

\section{Ground state cooling with sequential filtering}
\label{Sect_repeatF} We have seen that filtering is  limited by
the fact that it cannot correct defects that arise from holes in a
perfect MI phase. In order to circumvent this problem, we will now
consider a repeated application of filtering, which will clearly
profit from the increasing cooling efficiency as temperature is
decreased. Our approach requires
to iterate the following sequence of operations: (i) we allow for
some tunnelling while the trap is adjusted adiabatically in order
to reach a central occupation of $\la n_o \ra \approx 1.5$; (ii)
we suppress tunnelling and perform the filtering operation $F_1$;
(iii) the trap is slightly opened so that the final distribution
resembles a thermal distribution of hard-core bosons. This way we
transfer "hot" particles from the borders to the center of the
trap, where they can be removed by subsequent filtering.

 We are interested in the
convergence of the entropy and temperature as a function of the
number of iterations. However, the adiabatic process is very
difficult to treat both analytically and numerically. Therefore we
distinguish in the following between three different scenarios
that are based on specific assumptions.

{ \bf (i) Thermal equilibrium:}  We assume that the entropy is
conserved and that the system stays in thermal equilibrium
throughout the adiabatic process. Since this condition will in
general not be fulfilled for  closed, isolated quantum systems,
the following analysis can only provide a rather rough description
of the real situation. To be more precise, we start from  a
thermal state with initial parameters $\beta$, $b$ and $\mu$.
After filtering and adiabatic evolution one has a thermal state in
the no-tunnelling regime with new parameters $\beta'$, $b'$ and
$\mu'$.
As we have shown in the previous sections, thermal states in the
no-tunnelling regime can effectively be described in terms of two
fermionic components. This allows us to determine the new
parameters $\beta'$ and $b'$ by identifying: $S_\textrm{I}(\beta,
b)=S_\textrm{I}(\beta',b')+ S_\textrm{II}(\beta',b')$ and
$N_\textrm{I}(\beta, b)=N_\textrm{I}(\beta',b')+
N_\textrm{II}(\beta',b')$. The desired central filling $\la n_0
\ra=1.5$ fixes the chemical potentials to be $\mu=\mu'=U$. Using
expressions (\ref{S_I}) and (\ref{S_II}) one finds: \bea \beta'
U & = & \left(A+\sqrt{A^2+ 4 \beta U}\right)^2/4 , \label{beta'} \\
\frac{b'}{U} & = & \frac{b}{U}\left( 1+\frac{\eta_\textrm{II}}{2
\sqrt{\beta' U}}\right) \label{b'}, \eea with $A=(
\sigma_\textrm{II} \beta U /
\sigma_\textrm{I}-\eta_\textrm{II})/2$. After a second filtering
operation the entropy per particle is thus given by: \be
\label{s_2} s_2= \sigma_\textrm{I} \ \frac{1}{\beta' U}. \ee Since
our analysis only holds  in the limit $\beta U \gg 1$, one can
simplify the above expressions to: $\beta' U \approx
(\sigma_\textrm{II} \beta U /(2 \sigma_\textrm{I}))^2$ and
$b'\approx b$. This allows us to establish a simple recursion
relation for the entropy per particle $s_n$ after the $n$-th
filter operation: \be \label{sn} s_n \approx \frac{4
\sigma_\textrm{I}}{\sigma_\textrm{II}^2} \ s_{n-1}^2.\ee Since the
limit $\beta U \gg 1$  implies $s < 1$, one finds that the entropy
per particle converges extremely fast to zero.

\begin{figure}[h]
\centering \epsfig{file=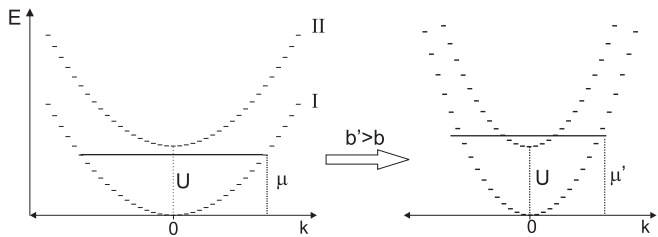,width=1\linewidth}
\caption{Effective description of thermal states in the
no-tunnelling limit in terms of non-interacting fermions occupying
two energy bands. The dispersion relations are $\eps_\textrm{I}=b
k^2$ and $\eps_\textrm{II}=b k^2  +U$, where $k$ denotes the
lattice site and $U$ is the interaction energy. Increasing the
harmonic trap strength from $b$ to $b'$ increases the chemical
potential to $\mu'$ so that the population of the upper band
becomes energetically favorable. In the bosonic picture this
process corresponds to the formation of doubly occupied sites.}
\label{fig_bands_2}
\end{figure}

{\bf  (ii) Quantum evolution:} Let us now study  a more realistic
situation. To this end we resort to an effective description of
the quantum dynamics in terms of two coupled Fermi bands [Appendix
A ]:
\begin{eqnarray} \tilde{H} &=& \sum_k \  [ b k^2 c_k^\dagger c_k +
(b k^2 + U) d_k^\dagger d_k
 \nonumber \\
 &-& J \ (c^\dagger_k c_{k+1}+ \textrm{h.c.}) \nonumber \\
&-& \sqrt{2} J \ (c^\dagger_k d_{k+1}+ d^\dagger_k c_{k+1}+
\textrm{h.c.} )\nonumber \\
 &-& 2 J \  (d^\dagger_k d_{k+1} + \textrm{h.c.})
]. \label{H_fermi}
\end{eqnarray}
Here, operators $c_k, c_k^\dagger$ refer to energy band I and
$d_k,
 d_k^\dagger$ to band II, which is shifted from the lower band by the
 amount of the interaction energy $U$ (see Sect. \ref{Init} and Fig. \ref{fig_bands_2}).
 This treatment is self-consistent as long as the
probability of finding a particle-hole pair is negligible,
i.e. $\la c_k c_k^\dagger d_k^\dagger d_k \ra \approx 0$.
In Sect. \ref{Init} we have shown analytically that this is
indeed fulfilled for thermal states in the no-tunnelling
regime and for low temperatures ($\beta U \gg 1$). We have
checked that it also holds for thermal states at finite
hopping rate $J$, provided that $J$ is not too big ($J
\lesssim 0.5 ~U$).  Since the Hamiltonian (\ref{H_fermi})
is quadratic, we can study the complete protocol in terms
of a single-particle picture. This can easily be done, if
one further assumes that no level crossings occur in the
course of the adiabatic evolution. Then,  the state at any
time $t$ can be computed by populating the single-particle
energies of $\tilde{H} (t)$ according to the initial
probabilities (after filtering) in energetically increasing
order. This method is illustrated in more detail in Fig.
\ref{fig_bands_2}.

 After the initial filtering step only states in the lowest energy band are occupied. The
 occupation probability is given by the fermionic  distribution $f_k(b, \beta, \mu)$ (\ref{nf}).
Increasing the trap strength to an appropriate value $b'>b$ in the
course of the adiabatic process makes it energetically favorable
to occupy also the second band. We find the state after returning
to the no-tunnelling regime, $\rho'$,  by populating the energy
levels, corresponding to the new trap strength $b'$,  in
energetically increasing order according to the initial
probabilities $f_k(b, \beta, \mu)$. At this point we distinguish
between two further scenarios: {\bf (ii.a)} The state $\rho'$ is
mapped to a thermal state in the usual way by accounting for
energy and particle number conservation [Sect. \ref{Init}]. This
way we can quantify the amount of "heating" resulting from the
fact that the system is not in thermal equilibrium at the end of
the adiabatic process due to the different structure of the energy
spectrum. {\bf (ii.b)} The next filtering operation acts directly
on the time evolved state $\rho'$.

\begin{figure}[h]
\centering \epsfig{file=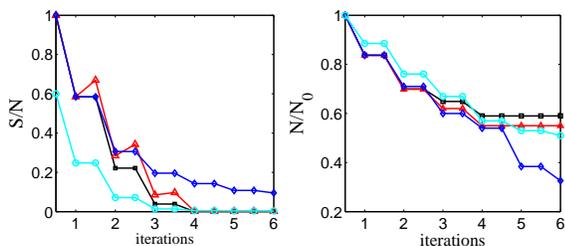,width=\linewidth}
\caption{(Color online) Entropy per particle $S/N$ (left) and
normalized number of particles $N/N_0$ (right) as a function of
the number of filtering iterations for initially $N_0=100$
particles. As discussed in the text, we distinguish between the
scenarios (i) (black line), (ii.a) (red line) and (ii.b) (blue and
cyan).} \label{fig_sequ_filter}
\end{figure}
Our results for all three cases are summarized in \mbox{Fig.
\ref{fig_sequ_filter}}. We have computed numerically exact the
entropy per particle as a function of the number of filtering
cycles. Starting with an initial entropy $s_0=1$, scenarios (i)
and (ii.a)  predict that an entropy value close to zero can be
obtained after only four iterations of the protocol \footnote{Note
that for thermal states at very low temperatures the entropy is
concentrated in only a few particles. Hence, finite size effects
become important and the minimal attainable entropy per particle
depends very sensitively on the strength of the harmonic
confinement.}. According to our underlying assumptions the system
is in thermal equilibrium after each iteration of the protocol. In
scenario (ii.b) the final  entropy saturates at a finite value and
the system is not in perfect thermal equilibrium. However, the
final state  still resembles a thermal state of hard-core bosons
in a harmonic trap.

These results imply that sequential filtering can clearly profit
from equilibration. The reason is that equilibration reduces the
defect probability in the center of the lower band and transfers
entropy to the upper band, where it can be removed subsequently.
This process in combination with the increasingly high cooling
efficiency of filtering at low temperatures can easily compensate
the heating induced by  the adiabatic quantum evolution. From our
data we can  deduce that this heating corresponds to an entropy
increase of around 20 \% \cite{com_heating}. Without equilibration
sequential filtering becomes very  inefficient after the fourth
iteration, which is also reflected in the  excessive particle loss
[Fig. \ref{fig_sequ_filter}]. The minimal attainable entropy is
determined by the initial defect (hole) probability in the center
of the trap. Starting from a much colder state, which exhibits
almost unit filling in the center of the lower band, therefore
yields  a final state very close to the ground state [Fig.
\ref{fig_sequ_filter}].

Remember that scenario (ii.b) is based on the assumption
that no level crossings occur during the evolution. From
our numerical analysis of the energy spectrum of
(\ref{H_fermi}) we know, however, that level crossings can
indeed appear (see also \cite{Menotti,Altman}). The reason
is the vanishing small spatial overlap between single
particle states at the border of the lower band and the
center of the upper band. This has the following
consequences for our previous analysis: For rather small
particle numbers ($N\lesssim 15$) level crossings are rare
and inter-band coupling occurs  already for hopping rates,
which are well within the range of validity of our
single-particle description. We therefore expect that our
predictions, as depicted in Fig. \ref{fig_sequ_filter}, are
reasonable.  For larger systems one has to tune the
tunnelling rate deep into the superfluid regime $J\gtrsim
0.5 ~U$ in order to couple the two bands and to form doubly
occupied sites. However, this regime is no longer
accessible within our fermionic model (\ref{H_fermi}). It
remains to be investigated to what extent this will alter
our predictions for the  cooling efficiency of sequential
filtering.


\section{Algorithmic  ground state cooling} \label{Sect_gscool}
\subsection{The protocol}
We now propose  a second cooling scheme, which we call algorithmic
cooling, because it is inspired by quantum computation. As before
the goal is to remove high energy excitations at the borders of
the atomic cloud, which have been left after filtering. In
contrast to sequential filtering we now restrict ourselves to a
sequence of quantum operations that operate solely in the
no-tunnelling regime. The central idea is to make use of
spin-dependent lattices. A part of the atomic cloud can then act
as a "pointer" in order to address lattice sites which contain
"hot" particles. In this sense the scheme is similar to
evaporative cooling, with the difference that an atomic cloud
takes the role of the rf-knife. Another remarkable feature of the
protocol is that the pointer is very inaccurate in the beginning
(due to some inherent translational uncertainty in the system),
but becomes sharper and sharper in the course of the protocol.
\begin{figure}[h]
\centering \epsfig{file=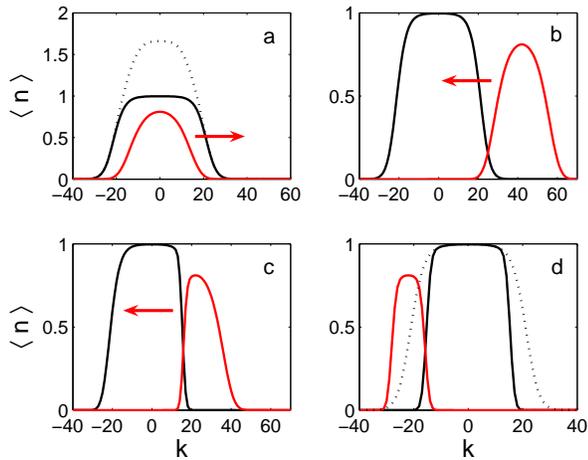,width=\linewidth}
\caption{(Color online) Illustration of the protocol. The state is
initialized with the filter operation $F_2$. (a) Particles from
doubly occupied sites are transferred to state $|b\ra$ (red) using
operation $U_{2,0}^{1,1}$. (b) The lattice  $|b\ra$ has been
shifted $2 k_\eps$ sites to the right, so that the two
distributions barely overlap. (c) Density distribution after $k_s$
lattice shifts. After each shift doubly occupied sites have been
emptied. Afterwards lattice $|b\ra$ is shifted $4 k_\eps - k_s$
sites to the left and an analogous filter sequence is applied. (d)
The final distribution of atoms in state $|a\ra$ is sharper
compared to the initial distribution (dotted). Numerical
parameters: $N_i=65$, $s_i=1$, $U/b=300$, $k_\eps$=21, $k_s=20$,
$N_f=30.2$, $s_f=0.31$ (after equilibration).} \label{fig_gs_cool}
\end{figure}

The steps of this algorithmic protocol are: (i) We begin with a
thermal equilibrium cloud with two or less atoms per site, all in
internal state $|a\ra$, and without hopping. This can be ensured
with a filtering operation $F_2$. (ii) We next split the particle
distribution into two, with an operation $U_{2,0}^{1,1}$
[Fig.~\ref{fig_gs_cool}a]. (iii) The two clouds are shifted away
from each other until they barely overlap. Then we begin moving
the clouds one against each other, emptying in this process all
doubly occupied sites. This sequence sharpens the density
distribution of both clouds. It is iterated for a small number of
steps, of order $\Delta$ (\ref{Delta}) [Fig.~\ref{fig_gs_cool}c].
(iv) The atoms of type $|b\ra$ are moved again to the other side
of the lattice and a process similar to (iii) is repeated
[Fig.~\ref{fig_gs_cool}d].  (v) Remaining atoms in state $|b\ra$
can now be removed.

The final particle distribution cannot be made arbitrarily sharp
[Fig.~\ref{fig_gs_cool}d], due to the particle number uncertainty
in the tails of the distribution. In the following we will
consider this argument more rigorously and develop a theoretical
description of the protocol.

\subsection{Theoretical description}
For the sake of simplicity we consider a slightly  modified
version of the protocol. The particle distributions $|a\ra$ and
$|b\ra$ are now two identical but independent distributions of
hard-core bosons of the form (\ref{nf}) \cite{Comment_theory}.
\begin{figure}[h]
\centering \epsfig{file=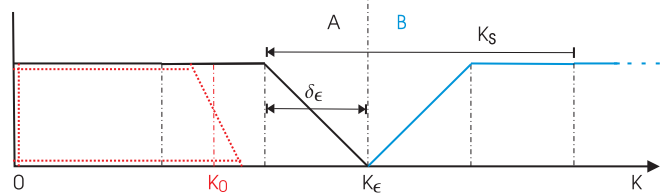,width=\linewidth}
\caption{(Color online) Schematic description of the initial state
for lattice sites $k\geq 0$: Two identical density distributions
for hard-core bosons, belonging to different species A (black) and
B (blue), are shifted $2 k_\eps$ lattice sites apart. The region
of non-integer filling has width $\delta_\eps$. In the course of
the protocol the lattice of species B is shifted $k_s=3
\delta_\eps$ sites to the left.} \label{fig_theory_schematic}
\end{figure}
The lattice $|b \ra$ is shifted $2 k_\eps$ sites to the right. For
given $\eps$ the value $k_\eps=\sqrt{\mu/b}\sqrt{1- \ln \eps /
\beta \mu}$ is chosen such that for atoms $|a \ra$ it holds $\la
n^a_{k_\eps} \ra= \eps$. This initial situation is depicted
schematically in Fig. \ref{fig_theory_schematic}. The cutoff
$\eps$ defines also the width of the region with non-integer
filling: \be \label{delta_eps} \delta_\eps = \sqrt{\mu/b} \left(
\sqrt{1- \ln \eps /\beta \mu} - \sqrt{1 + \ln \eps /\beta \mu}
\right). \ee

We analyze first a protocol that involves  $k_s=3 \delta_\eps$
lattice shifts and after each shift doubly occupied sites are
emptied. Our goal is to compute the final shape of the density
profile of atoms in state $|a\ra$ (red line in Fig.
\ref{fig_theory_schematic}). It is sufficient to consider only the
reduced density matrices $\hat \rho_a$ and $\hat \rho_b$, which
cover the range $k \in [k_\eps -2 \delta_\eps; k_\eps]$ and $k \in
[k_\eps;k_\eps +2 \delta_\eps]$, respectively. These density
matrices can be written in terms of convex sums over particle
number subspaces: \bea \hat \rho_a & = &
\sum_{N_a=0}^{2 \delta_\eps} p_a (N_a) \hat \rho_a(N_a), \\
\hat \rho_b & = & \sum_{N_b=0}^{2 \delta_\eps} p_b (N_b) \hat
\rho_b(N_b). \eea

The further discussion is based on the following central
observation. If a state $\hat \rho_a (N_a)$ interacts with a state
$\hat \rho_a (N_b)$ then our protocol produces a perfect MI state
$\hat \rho'_a(N'_a)$ composed of $N'_a=(N_a-N_b)/2$ particles. The
factor $1/2$ arises from the fact that $k_s$ lattice shifts remove
at most $k_s/2$ particles from distribution $|a\ra$. Note that
this relation also allows for negative particle numbers, because
$N'_a$ merely counts the number of particles on the right hand
side of the reference point $k_0=k_\eps-3/2 \delta_\eps$. The
final density matrix after tracing out particles in $|b\ra$ can
then be written as a convex sum over nearly perfect (up to the
cutoff error $\eps$) MI states \be \label{rho'} \hat \rho'_a =
\sum_{N'_a=-\delta_\eps/2}^{\delta_\eps/2} p'_a (N'_a) \hat
\rho'_a(N'_a), \ee with probabilities

\be \label{pN'} p'_a(N'_a) \simeq 2  \sum_{N_b=\delta_\eps}^{2
\delta_\eps} p_a(2 N'_a+N_b) p_b(N_b). \ee The factor two is due
to the fact that states with $N_a-N_b=2 M$ and $N_a-N_b=2M+1$ are
collapsed on the same MI state with $N_a'=M$. Since Lyapounov's
condition \cite{CLT} holds in our system, we can make use of the
generalized central limit theorem and approximate $p_a(N)=p_b(N)$
by a Gaussian distribution with variance $\sigma^2=\delta
N^2=\Delta /4= 1/(2\beta \sqrt{b
 \mu})$. Evaluation of Eq. (\ref{pN'}) then yields a Gaussian
 distribution with variance $\sigma'^2=\sigma^2/2$. Since MI
 states do not contain holes, one can infer the final density
 distribution $\la n^a_k \ra'$ directly from $p_a'(N)$ by simple integration.
This distribution can then be approximated  by the (linearized)
thermal distribution: \be \label{n'} \la n_k \ra' \simeq
\frac{1}{1+e^{4 (k-k_0) / \Delta'}}. \ee The new effective tail
width $\Delta'=\sqrt{\Delta \pi }/2$ of the distribution is
roughly the square root of the original \mbox{width $\Delta$
(\ref{Delta})}. This effect leads to cooling, which we will  now
quantify in terms of the entropy.

When applying similar reasoning also to the left side of
distribution $|a\ra$ one ends up with a mixture of MI states,
which differ by their length \emph{and} lateral position. This
results in an extremely small entropy of the order
$S_{\textrm{MI}}\sim \log_2{\Delta}$. However, this final state is
typically far from thermal equilibrium. In order to account for a
possible increase of entropy by equilibration, we have to compute
the entropy of a thermal state, which has the same energy and
particle number expectation values as the final state. In our case
this is equivalent to computing the entropy directly from the
density distribution (\ref{n'}): \be \label{S'} S'\approx
\sigma_\textrm{I}~ \Delta' =  \frac{\sigma_\textrm{I} \sqrt{\pi}}{
\sqrt{2}} \frac{1}{(\beta^2 b \mu)^{1/4}}
 = \frac{\sqrt{\pi}}{2}  \frac{\sqrt{\beta \mu}}{\sqrt{N}} \ S , \ee
where $N= N_\textrm{I}$ and $S= S_\textrm{I}$ (\ref{S_I})
correspond to the expectation values after the initial filtering
operation. The final particle number is given by: $ N' \simeq 2
k_0 \approx N(1+\ln \eps / \beta \mu)$.


A significant improvement can be made by shifting the clouds only
$k_s=2 \delta_\eps$ sites. This prevents inefficient particle
loss, which has  been included  in our previous analysis in order
make the treatment exact. With this variant the final particle
number increases to $N''=2~(k_\eps- \delta_\eps)$, while in good
approximation the final entropy is still given by $S'$. Hence, the
final entropy per particle can be lowered to: \be
 \label{s'} s'= \frac{S'}{N''} \approx \frac{\sqrt{\pi}}{2} ~
\frac{\sqrt{\beta \mu}}{1+ \frac{\ln \eps}{2 \beta \mu}} ~
\frac{1}{\sqrt{N}} \ s.  \ee
 This expression, which holds strictly only in the limit $\beta U \gg 1$, shows that
 for fixed $N$ the ratio $s'/s$ becomes smaller
 at higher temperatures. Even more important, for fixed $\beta U$,
 the entropy per particle $s'$ decreases with $1/\sqrt{N}$ as the
 initial  number of particles in the system increases.

Finally, let us remark that the final entropy can be further
reduced, when the protocol is repeated with two independent states
of the form (\ref{rho'}). In practice, this could be achieved with
an ensemble of non-interacting atomic species in different
internal levels. According to Eq. (\ref{pN'}), each further
iteration of the protocol decreases the total entropy by a factor
$1/\sqrt{2}$.

\section{Discussion of results}\label{Sect_discussion}
\begin{figure}[h]
\centering
\epsfig{file=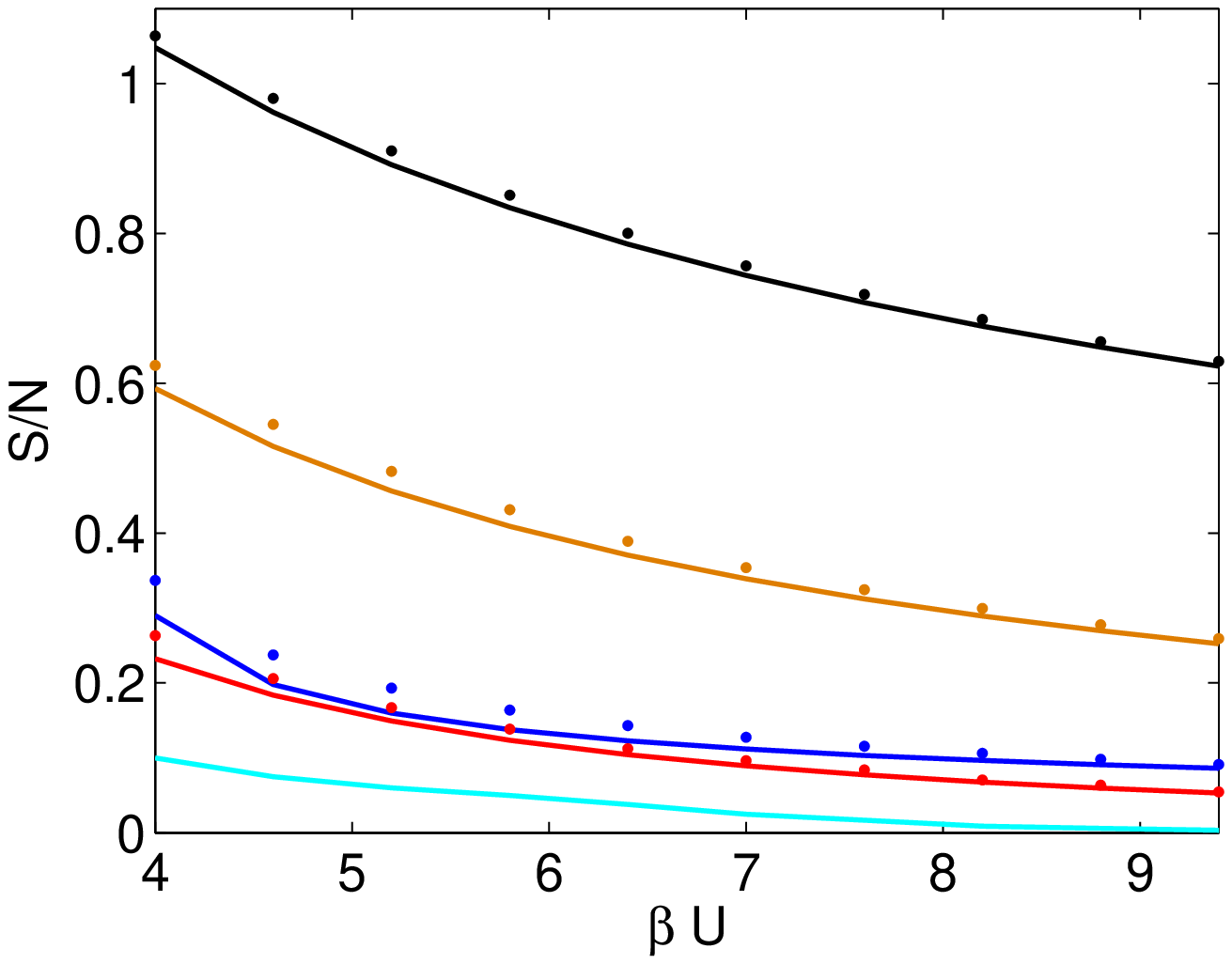,width=0.8\linewidth,angle=0}
\epsfig{file=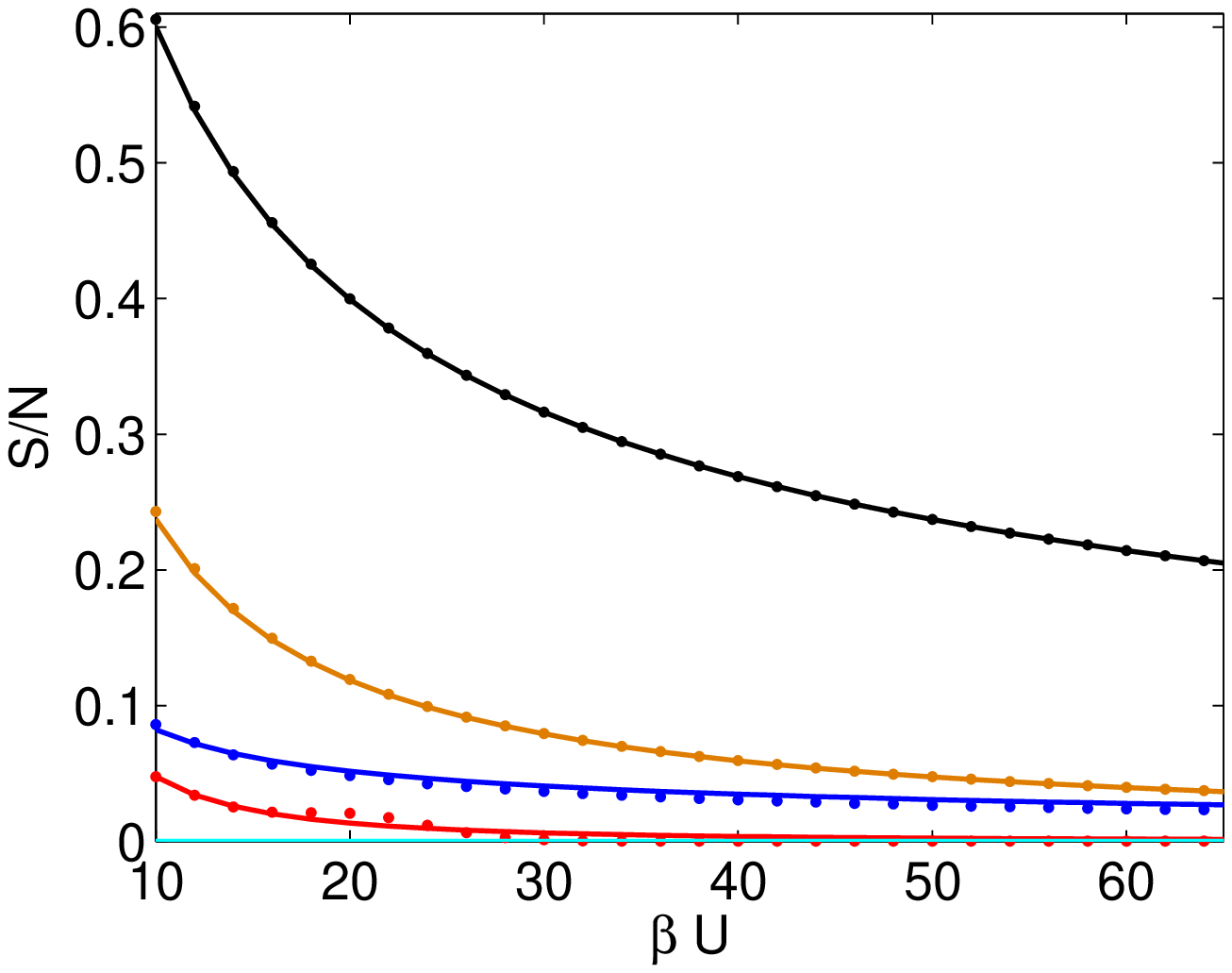,width=0.8\linewidth,angle=0}
\caption{(Color online) Analytical (solid) and numerical (dots)
results for the entropy per particle $s=S/N$ versus the inverse
temperature $\beta U$ for fixed trap strength $b/U=1/2500$ and
fixed $\mu=U$. We plot the initial value (black line) and the
values after filtering with $F_1$ (Eq. (\ref{s_f}), brown line).
This is compared to more elaborate cooling protocols: entropy
after two iterations of sequential filtering including
equilibration (Eq. (\ref{s_2}), red line), minimal entropy after
sequential filtering without equilibration (cyan), and entropy
after algorithmic cooling (Eq. (\ref{s'}), blue line). For the
algorithmic protocol we have chosen $\eps=0.03$ and
$k_s=2~\delta_\eps$.  Note that this protocol creates classical
correlations between lattice sites. The numerics are therefore
based on a representation of classical density matrices in terms
of matrix-product states \cite{Cool2}.} \label{fig_checktheory}
\end{figure}

Let us now discuss and compare our previous results. In
particular, we are interested in checking the range of validity of
our analytical results by comparison with exact numerical
calculations. To this end, we fix two parameters, $b/U$ and
$\mu/U$, and compute the entropy per particle as a function of the
inverse temperature $\beta U$ [Fig.~\ref{fig_checktheory}].

Our results can be summarized as follows. Firstly, we find that
our theoretical description is very accurate in the low
temperature limit $\beta U \gg 1$, and even holds in the
(relatively) high temperature range $\beta U \gtrsim 1$. Secondly,
our algorithmic protocol outperforms filtering considerably,
especially in the high temperature range. Finally,
 based on the  assumptions that underlie our calculations, subsequent  filtering is typically
 superior to algorithmic cooling with respect to the minimal achievable entropy.

We now discuss advantages, experimental requirements and time
scales of our  cooling protocols.
\subsection{Sequential filtering}
(i) \emph{Advantages}: If one combines filtering with
equilibration then the entropy per particle converges very fast to
zero with the number of filter steps. The minimal value is only
limited by finite size effects. Without equilibration, the minimal
entropy is limited by the finite probability of finding a hole in
the central MI phase of the initial distribution. Furthermore,
sequential filtering naturally allows for cooling in a 3D setup,
because it preserves spherical symmetry. Note that filtering, and
hence sequential filtering, can also be applied to fermionic atoms
in an optical lattice \cite{RZ03}.
\\
(ii) \emph{Requirements and limitations}: The repeated creation of
doubly occupied sites in the center of the trap requires precise
control of the harmonic confinement over a wide range of values.
In addition, non-adiabatic changes of lattice and external
potentials might induce heating, which could reduce the cooling
efficiency
considerably. \\
(iii) \emph{Time scales}: The limiting factor here is not
filtering but the adiabaticity criterion for changes of the
hopping rate and the harmonic confinement. We have shown that
after filtering the density distribution can be identified with a
thermal distribution of spin-less fermions. Hence it is possible
to find estimates for adiabatic evolution times based on single
particle eigenfunctions, as calculated in \cite{Menotti}. In the
MI regime particle transfer from the borders to the center of the
trap is very unlikely, because the eigenfunctions of the upper and
lower Fermi band do not overlap. Therefore, we propose as a first
step to decrease the lattice potential at fixed harmonic
confinement until eigenfunctions start overlapping. This process
can still be treated within a fermionic (or Tonks gas) picture,
since only the lower band is populated [Fig. \ref{fig_bands_2}].
The average energy spacing around the Fermi level is
$\overline{\delta E} = b N \approx 2~ U/N$ in the no-tunnelling
regime, and stays roughly constant when passing over to the
tunnelling regime \cite{Menotti}. As a consequence, the lattice
potential should be varied on a time scale $T \gg h/
\overline{\delta E} \approx 10$ ms for $N=50$ and $h/U=396~ \mu$s
\cite{WB04}. For the second process, which involves the change of
the harmonic potential at fixed hopping rate, it is more difficult
to make analytic predictions for adiabatic time scales, since our
single-particle description (\ref{H_fermi}) is  no longer
justified in general. One can, however, obtain a lower bound by
considering the energy spectrum after returning to the
no-tunnelling regime [Fig. \ref{fig_bands_2}]. The average energy
spacing around the Fermi level is now dominated by the energy
spacing at the bottom of the upper band: $\overline{\delta E'}
=\sqrt{b'/\beta}/2 \approx U/(N \sqrt{\beta U})$. This implies
adiabatic evolution times which are a factor $\sqrt{\beta U}/2$
larger than for the first process. We have also verified the whole
process numerically, using the matrix-product state representation
of mixed states \cite{DMRGmixed}. For initially $N=11$ particles
we find adiabatic evolution times of the order $T \sim 30 ~\hbar /
U$ for the first process, which is consistent with our analytical
estimate. The second process is more time consuming with $T
\gtrsim 120~ \hbar/U$.
\subsection{ Algorithmic cooling}
(i) \emph{Advantages}:
 The algorithmic protocol  operates
solely in the no-tunnelling regime. Adiabatic changes of the
lattice and/or the harmonic potential, which are time consuming
and might induce heating,  are therefore not required. Moreover it
does not demand arbitrary control over the harmonic confinement.
The correct initial conditions  can always be generated by the
filter operation $F_1$. The protocol is more efficient in the high
temperature range and for large particle numbers. Additionally to
ground state cooling, the algorithm can be used to generate an
ensemble of nearly perfect quantum registers for quantum
computation. This state, which can  also be  considered as an
ensemble of possible ground states in the uniform system, might
already be sufficient for quantum simulation of certain spin
Hamiltonians. Finally note that this
protocol can naturally be applied also to fermionic systems. \\
(ii) \emph{Requirements and limitations}: The heart of the
protocol is the existence and control of spin-dependent lattices.
Moreover, the algorithm is  explicitly  designed for cooling in
one spatial dimension. Generalizations to higher dimensions are
possible, but will typically not preserve the spherical symmetry
of the initial density distribution. Moreover, one should keep in
mind that the
final states are typically far from thermal equilibrium.\\
(iii) \emph{Time scales}: Adiabatic lattice shifts can be
performed very fast on a  time scale determined by the on-site
trapping frequency $\nu \simeq 30$ kHz. The limiting factor is the
number of filter operations, which is of the order $\delta_\eps$
(\ref{delta_eps}). Under realistic conditions this can amount to
50 operations. Filter operation  times based on the adiabatic
scheme \cite{RZ03} are of the order $T_{F}\sim 200 ~\hbar/U$. With
$h/U=396~ \mu$s \cite{WB04} one finds a total operation time
$T\sim 630 ~$ ms, which is comparable with the typical particle
life time in present setups using  spin-dependent lattices
\cite{Bloch03}. We have studied this problem with an alternative
implementation of filtering [Sect. \ref{Sect_ultrafast}], which
allows to reduce operation times by a factor of $\sim 15$, and
hence makes algorithmic cooling feasible in current experimental
setups.

\section{Ultra-fast filtering scheme} \label{Sect_ultrafast}

We now propose an ultra-fast, coherent implementation of
filtering, which is based on optimal laser control. We restrict
our discussion to the filtering operation $F_1$ which is most
relevant for our cooling protocols discussed above. We consider
atoms in a particular internal level, $|a\ra$, which are coupled
to a second internal level, $|b\ra$, via a Raman transition with
Rabi frequency $\Omega(t)$. In contrast to the adiabatic scheme
\cite{RZ03} we consider constant detuning, but vary  $\Omega(t)$
in time. The Hamiltonian for a single lattice site reads \bea \hat
H &=& \frac{U_a}{2} \hat n_a(\hat n_a -1)+ U_{ab} \hat n_a \hat
n_b+\frac{U_b}{2} \hat n_b(\hat n_b -1) \nonumber\\
& & - (\Omega(t) \hat a^\dagger \hat b+ \Omega^*(t) \hat b^\dagger
\hat a), \label{H_opt}\eea where $U_a$, $U_b$ and $U_{ab}$ denote
the on-site interaction energies, according to the different
internal states. Note that $\Omega(t)$ can be complex, thus
allowing for time-dependent phases. Our goal is to populate state
$|a\ra$ with exactly one particle per site which can be expressed
by the unitary operation $U_0: |N,0\ra \rightarrow |1,N-1\ra, \
\forall \ N \in \{1,2,\ldots, N_{max} \}$. In order to do this, we
control the time-dependence of $\Omega(t)$ coherently and in an
optimal way. To be more precise, we optimize a sequence of $M$
rectangular shaped pulses of equal length: \be \label{Omega_pulse}
\Omega(t)=\sum_{l=1}^M \Omega_l \times
[\theta(t-t_l)-\theta(t-t_{l+1})]. \ee
After time $T$ the system has evolved according to the unitary
operator $U(T)$. We want to minimize the deviation of $U(T)$ from
the desired operation $U_0$, which we quantify by the infidelity $
\epsilon(T)=\sum_{N=1}^{N_{max}} \epsilon_N$, where $\epsilon_N =
1-|\la 1,N-1 |U(T)| N,0 \ra|$. Since we allow for complex Rabi
frequencies, $\epsilon(T)$ is a function of $2 M$ parameters $\{
\Omega_l, \Omega_l^* \}$ with $l=1, \ldots, M$. For given $M$ and
time $T$ we optimize the cost function $\epsilon(T)$ numerically
using the Quasi-Newton method with a mixed quadratic and cubic line
search procedure. This is repeated for different times $T$, while
keeping the number of pulses $M$ constant. We then increase $M$ and
repeat the whole procedure in order to check convergence of our
results.
\begin{figure}[h]
\centering
\epsfig{file=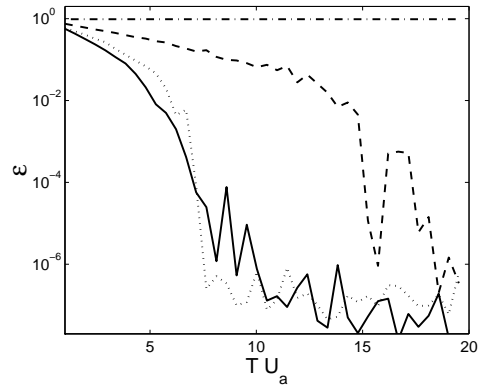,width=0.80\linewidth,angle=0}
\caption{Operation error $\epsilon$ vs. time $t U_a$ for different
  interaction energies: $U_b=0.2 U_a$, $U_{ab}=0.2 U_a$, $\delta=0.8$
  (solid); $U_b=U_a$, $U_{ab}=0.6 U_a$, $\delta=0.8$ (dashed);
  $U_b=U_a$, $U_{ab}=0.2 U_a$, $\delta=1.6$ (dotted);
  $U_b=U_a=U_{ab}$, $\delta=0$ (dashed-dotted). We optimize a sequence
  of $M=10$ pulses (\ref{Omega_pulse}).
}
\label{fig_opt_control}
\end{figure}
In Fig.~\ref{fig_opt_control} we plot the minimal error $\eps(T)$
for different interaction strengths. Already for our simple control
scheme we obtain very small errors, $\epsilon\sim 10^{-4}$, for a
time $T\approx 7/U_a$ and interactions $U_{ab}=U_b=0.2U_a$. In
comparison, the adiabatic scheme \cite{RZ03} would require $T\approx
150/U_a$ for the same set of parameters. However, in deep contrast
to \cite{RZ03} our method works for a very broad range of
interaction energies and in addition allows for very high
fidelities.

It is important to remark that the operation time increases
 for  small interaction anisotropies
$\delta=|U_a+U_b- 2 U_{ab}|/U_a$ and for  $\delta=0$ our method
fails.  In the special case $U_a=U_b=U_{ab}$ this follows from the
fact that in the Hamiltonian (\ref{H_opt}) the interaction part
commutes with the coupling part. These problems can be solved,
either by displacing the lattices that trap atoms $|a\ra$ and
$|b\ra$ and thereby reducing the effective interaction, $U_{ab}$,
or by performing more elaborate controls than the one from Eq.
(\ref{Omega_pulse}).

\section{Conclusion}
We have given a detailed analytical analysis of filtering in the
no-tunnelling regime and in the presence of a harmonic trap. We
have found that the residual entropy after filtering is localized
at the borders of the trap, quite similar as in fermionic systems.
Inspired by this result we have proposed two protocols that aim at
removing particles from the borders and thus lead to cooling. One
scheme transfers particles from the borders to the center of the
trap, where they can be removed by subsequent filtering. In the
other, algorithmic protocol particles from the system itself take
the role of the rf-knife in evaporative cooling and remove
directly particles at the borders. We have quantified the cooling
efficiency of these protocols analytically in terms of the initial
and final entropy per particle. To this end, we have also
developed an effective description of the single--band
Bose-Hubbard model
 in terms of two species of non-interacting fermions.

A special virtue of our schemes is that they rely on general
concepts which can easily be adapted to different experimental
situations.  For instance,  our protocols can easily be extended
to fermionic systems (for details see \cite{Cool2}). Moreover, the
algorithmic protocol can be improved considerably by the use of an
ensemble of non-interacting atomic species in different internal
states. In this context one should also keep in mind that a 3D
lattice structure offers a large variety of possibilities, which
have not been fully explored yet.

Since we have  identified the limitations of filtering,  one can
immediately think of alternative or supporting cooling schemes.
For instance, one might use ring shaped lasers beams to remove
particles at the borders of a 3D lattice. Equivalently, one can
use a transition, which is resonant only for atoms with
appropriate potential energy \cite{Cool2}. Although these methods
do not allow to address individual lattice sites, they might still
be useful as preliminary cooling steps for the protocols proposed
in this article.

We believe that the ideas introduced in this article greatly
enhance the potential of optical lattice setups for future
applications and might pave the way to the experimental
realization of quantum simulation and adiabatic quantum
computation in this system. We also hope that  our analytical
analysis of the virtues and limitations of current proposals,
especially filtering, might trigger further research in the
direction of ground state cooling in optical lattices.

\section{Acknowledgements}
This work was supported in part by EU IST projects (RESQ and
QUPRODIS), the DFG, and the Kompetenznetzwerk
``Quanteninformationsverarbeitung'' der Bayerischen
Staatsregierung.
\begin{appendix}
\section{Mapping to two-band fermionic model}
We start from the single-band Bose-Hubbard Hamiltonian (\ref{BHM})
and restrict the occupation numbers at each lattice site $k$ to
$n_k  \in \{ 0, 1, 2 \} $. In this truncated basis the Hamiltonian
reads:
\begin{eqnarray}
H &=& \sum_k \  [ b k^2  |1 \ra_k \la 1| + 2 (b k^2 + U) |2 \ra_k \la 2| \nonumber\\
&-& J \ (|0\ra_k |1\ra_{k+1} \la 1|_k \la 0|_{k+1} + \textrm{h.c.}) \nonumber\\
&-& \sqrt{2} J \ (|0\ra_k |2\ra_{k+1} \la 1|_k \la 1|_{k+1}+
\textrm{h.c.})\nonumber \\
&-& \sqrt{2} J \ (|2\ra_k |0\ra_{k+1} \la 1|_k \la 1|_{k+1}+ \textrm{h.c.}) \nonumber\\
&-& 2 J \ (|2\ra_k |1\ra_{k+1} \la 1|_k \la 2|_{k+1} +
\textrm{h.c.}) ].
\end{eqnarray}
One can now embed the three dimensional single site Hilbert space
$\mathcal{H}_B=\mathbb{C}^3$ into the composite Hilbert space
$\mathcal{H}_F=\mathbb{C}^2 \otimes \mathbb{C}^2$ of two species
of hard-core bosons by applying the following mapping:
\begin{eqnarray}
|2 \ra = \frac{1}{\sqrt{2}} (a^\dagger)^2 |\textrm{vac} \ra
&\rightarrow& \tilde{c}^\dagger  \tilde{d}^\dagger |\textrm{vac}
\ra , \nonumber \\
|1 \ra = a^\dagger |\textrm{vac} \ra &\rightarrow&
\tilde{c}^\dagger |\textrm{vac} \ra .\nonumber \\
\end{eqnarray}
Note that singly occupied bosonic states are mapped exclusively to
the $\tilde c$--manifold, i.e. we omit the possibility of having
one particle in the the $\tilde d$--manifold and no particle in
the $\tilde c$--manifold on the same site. After transforming
hard-core bosons to fermions, $\tilde{c}, \tilde{d} \rightarrow c,
d$, via a Jordan-Wigner transformation one obtains the following
fermionic Hamiltonian:
\begin{eqnarray}
H &=& \sum_k \  [ b k^2 c_k^\dagger c_k d_k d_k^\dagger+ (b k^2 + U) c_k^\dagger c_k d_k^\dagger d_k
\nonumber \\
&-& J \ (c^\dagger_k c_{k+1} d_k d_k^\dagger d_{k+1} d_{k+1}^\dagger + \textrm{h.c.}) \nonumber \\
&-& \sqrt{2} J \ (c^\dagger_k d_{k+1} d_k d_k^\dagger
c_{k+1}^\dagger c_{k+1}+ \textrm{h.c.}) \nonumber \\
&-& \sqrt{2} J \ (d^\dagger_k c_{k+1} c_k c_k^\dagger  d_{k+1} d_{k+1}^\dagger+ \textrm{h.c.}) \nonumber \\
 &-& 2 J \ (d^\dagger_k d_{k+1} c_k c_k^\dagger c_{k+1} c_{k+1}^\dagger + \textrm{h.c.})
]
\end{eqnarray}
This Hamiltonian can also be written in the form $H=P^\dagger
\tilde H P$, where $\tilde{H}$ is the quadratic Hamiltonian
(\ref{H_fermi}) and $P$ denotes the projection on the subspace,
which is defined by $ c_k^\dagger  d_k =0$. This implies that
bosonic atoms in an optical lattice can effectively be described
in terms of the quadratic Hamiltonian (\ref{H_fermi}), given that
the probability of finding a particle-hole pair is negligible,
\mbox{i.e. $\la  c_k c_k^\dagger d_k^\dagger d_k \ra \approx 0$.}

Let us point out that similar fermionization procedures have been
discussed in \cite{PC05,TPC_unp}
\end{appendix}


\begin{thebibliography}{10}
\bibitem{comment_no-tunnel} The condition for the no-tunnelling
regime in the presence of a shallow harmonic trap is given
by $|J/(bN)|^2 \ll 1$, i.e. the hopping matrix element is
much smaller than the average energy spacing between single
particle states located at the borders of the trap. If
desired one can also demand $|J/b|^2 \ll 1$, which ensures
that single particle eigenfunctions even at the bottom of
the trap are localized on individual lattice wells.
\bibitem{com_optimal}  It has turned out that this choice, incidentally,
yields the best cooling result \cite{Cool2} for a wide range of
initial conditions. In practice, it will therefore be the most
relevant case.
\bibitem{com_heating} We have also performed multi-particle calculations
in the canonical ensemble, which predict an even lower value of
about 5\%.
\bibitem{Comment_theory} In practice, this can be achieved by applying $F_1$ to two
non-interacting bosonic clouds in different internal states.
\end{thebibliography}
\end{document}